\documentclass[lettersize,journal]{IEEEtran}
\usepackage{amsmath,amsfonts}
\usepackage{algorithmic}
\usepackage{algorithm}
\usepackage{array}
\usepackage[labelsep=period]{caption}
\usepackage{textcomp}
\usepackage{stfloats}
\usepackage{url}
\usepackage{verbatim}
\usepackage{graphicx}
\usepackage{cite}
\usepackage{amssymb,amsthm}
\usepackage[colorlinks=true, linkcolor=blue]{hyperref}
\usepackage[table]{xcolor}
\usepackage{bm}
\usepackage{multirow}
\usepackage{makecell}
\usepackage{pifont}
\usepackage{subcaption}

\newtheorem{assumption}{Assumption}
\begin{document}
\title{UAV-Assisted Cooperative Edge Inference for Low-Altitude Economy via MoE-based Hierarchical Deep Reinforcement Learning}
\author{Wenhao Zhuang,~\IEEEmembership{Student Member,~IEEE},
        Yuyi Mao,~\IEEEmembership{Senior Member,~IEEE},
        \\ Ivan Wang-Hei Ho,~\IEEEmembership{Senior Member,~IEEE}, and Xianghao Yu,~\IEEEmembership{Senior Member,~IEEE}

\thanks{
    This paper was presented in part at the 2025 IEEE Global Communications Conference (GLOBECOM)~\cite{zhuang2025gc}.~\emph{(Corresponding author: Yuyi Mao.)}

    W. Zhuang and Ivan Wang-Hei Ho are with the Department of Electrical and Electronic Engineering, The Hong Kong Polytechnic University, Hong Kong, China (e-mail: wenhao.zhuang@connect.polyu.hk; ivanwh.ho@polyu.edu.hk).

    Y. Mao is with the School of Computer Science and Engineering, Macau University of Science and Technology, Taipa, Macau (e-mail: yymao@must.edu.mo).

    X. Yu is with the Department of Electrical Engineering, City University of Hong Kong, Hong Kong, China (e-mail: alex.yu@cityu.edu.hk).
}

\vspace{-1em}
}

\maketitle

\begin{abstract}
The low-altitude economy (LAE) is reshaping the industrial landscape by deploying unmanned aerial vehicles~(UAVs) to facilitate a wide range of applications demanding flexible aerial mobility. Integrating edge artificial intelligence (AI) into LAE platforms creates a compelling paradigm where UAVs provide real-time AI-driven analysis while simultaneously executing their primary aerial mission duties. 
However, realizing this paradigm remains challenging due to the strict mission constraints imposed by these primary duties and the throughput bottlenecks of wireless links. To bridge this gap, we propose a UAV-assisted cooperative edge inference framework where UAVs execute mission-critical LAE duties, quantified by trajectory deviations from reference paths, while concurrently supporting ground devices via intermediate feature offloading. Within this framework, UAV trajectories, inference task offloading decisions, and feature compression ratios are jointly optimized to maximize the system performance. We cast this joint optimization task into a constrained partially observable Markov decision process (POMDP) framework. To efficiently solve it, we propose HDRL-MoE, a novel hierarchical deep reinforcement learning framework that decouples the optimization of slow-varying inference decisions from rapidly changing UAV trajectory control. Furthermore, HDRL-MoE integrates a mixture-of-experts (MoE) architecture, where a router network orchestrates discrete offloading decisions while expert networks independently optimize the feature compression ratios. 
Extensive simulations show that HDRL-MoE achieves significant inference accuracy gains over baselines and exhibits high scalability and efficiency through its MoE design.
\end{abstract}


\section{Introduction}
\IEEEPARstart{T}{he} low-altitude economy (LAE) is rapidly emerging as a transformative economic paradigm, driven by the proliferation of unmanned aerial vehicle (UAV) applications such as aerial delivery, infrastructure inspection, and precision agriculture~\cite{laewhitepaper2024,yixian2025lae}. 
To enable strategic decision-making, smart Internet-of-Things (IoT) devices, including sensors and cameras, are increasingly deployed across LAE ecosystems to provide comprehensive sensory data, facilitating critical functions such as surveillance, anomaly detection, and environment monitoring to support the LAE operations~\cite{iotanduav2022}. 
However, these devices are typically constrained in computing capability and energy budget, rendering timely local processing of complex tasks impractical. The emergence of edge artificial intelligence (AI) inference, which deploys deep neural network (DNN) models at the network edge, opens a promising avenue for LAE. Equipped with onboard computing hardware, UAVs can function as mobile aerial servers, hosting AI inference tasks offloaded from ground devices (GDs) and delivering real-time services~\cite{greenedgeai2024}. Moreover, the inference results can be used for adapting the operational strategies of LAE missions in response to system dynamics~\cite{lyu2026empowering}. 

Nevertheless, the synergy between LAE and edge AI imposes a non-trivial burden on communication. While the high mobility of UAVs can be leveraged to alleviate the channel path loss, it remains insufficient to simultaneously support high-throughput transmissions for many GDs across a wide coverage area.
Therefore, analogous to conventional edge AI systems, offloading raw sensory data of GDs to the UAV is infeasible in LAE scenarios~\cite{uavcommun2016}. 
Cooperative inference resolves this bottleneck by partitioning a DNN model between a GD and the UAV~\cite{zeng2021coedge}. 
As such, each GD only needs to run a lightweight device-side DNN partition locally to extract compact intermediate features. These features, generally with considerably smaller size compared to the raw data, are then offloaded for processing within the UAV-side DNN partition. 
Unfortunately, despite the benefits, the realization of cooperative inference in LAE scenarios introduces several critical challenges. Specifically, LAE missions and AI inference are governed by heterogeneous performance metrics, which necessitates multi-objective considerations. Their performance is further entangled by the UAV trajectory, which not only determines the channel conditions for inference task offloading but also the effectiveness of regular LAE missions.
Furthermore, the limited communication and computational capabilities on UAVs require strategic resource allocation among IoT devices with competing demands. To this end, we propose a novel framework that jointly optimizes UAV trajectory and cooperative inference offloading decisions to maximize inference accuracy while satisfying LAE mission requirements.

\subsection{Related Work}
Performing edge AI inference on resource-constrained wireless networks presents a fundamental conflict among computation, communication, and task-intelligence. To mitigate this tension, efficient cooperative edge inference frameworks have been extensively investigated. For instance, a cooperative DNN inference system was proposed in~\cite{zeng2021coedge}, which partitions inference workloads across heterogeneous edge devices.
The rate-distortion tradeoff of cooperative inference was formalized via the information bottleneck (IB) framework in~\cite{jiawei2021ib}, demonstrating that pruning task-irrelevant features can substantially reduce communication overhead while preserving inference accuracy. Beyond communication, the computation distortion introduced by feature quantization was further incorporated in~\cite{dingzhu2024multi}.
To improve the robustness of compressed edge models against data drift during deployment, ensemble learning was adopted in~\cite{zhou2025federated}, where multiple edge devices collaboratively aggregate model predictions under resource budgets and latency constraints. 

While the above works advance the framework of cooperative edge inference, efficiently orchestrating the AI inference tasks of multiple devices necessitates judicious allocation of wireless resources. A mixed-integer nonlinear program was formulated in~\cite{lixian2024optimal} to jointly optimize DNN split points and resource allocation for multi-user cooperative inference. To pursue energy-efficient edge inference in wireless sensing systems, both communication and computation were optimized in accordance with sensing quality in~\cite{jiaocheng2025energy}. Moreover, an event-triggered offloading mechanism was proposed in~\cite{zhou2025communication} to suppress redundant communication by transmitting intermediate features only when data variations exceed a prescribed threshold. Nonetheless, these studies assume static ground edge servers. Since channel conditions vary with UAVs' positions in LAE scenarios, the interplay between trajectory planning and resource allocation decisions merits particular~attention.

Early efforts on UAV-enabled wireless networks target at jointly optimizing trajectory and user scheduling to maximize communication performance~\cite{qingqing2018uav,kaitao2023throughput}, while subsequent works have integrated UAVs with broader functionalities such as data collection~\cite{lai2024data} and sensing~\cite{gaoyuan2025isac,xiaowen2026isac}.
However, supporting AI inference with UAV-enabled wireless networks calls for a renewed perspective, as inference accuracy also depends on the task relevance of transmitted features. This consideration gives rise to additional optimization dimensions, e.g., data selection and feature compression. 
Recently, an IB-based framework was developed in~\cite{zhengru2025task} to prune redundant multi-view features for cooperative visual navigation in LAE scenarios. To enable joint edge AI inference and wireless sensing,~\cite{yangfu2024ai} proposed a UAV-enabled system where the UAV processes compressed intermediate features offloaded from GDs while performing target tracking.
Nevertheless, the feature compression strategy relies on offline statistical importance metrics rather than inference difficulty of individual data. In~\cite{jingfeng2025dynamic}, multi-device features were aggregated via over-the-air computation with joint UAV trajectory optimization and power allocation, yet this work only addresses inference performance.

The seamless integration of UAV-assisted cooperative edge inference under practical LAE constraints poses three critical challenges not fully addressed by existing works, as summarized in Table~\ref{tab:comparison}. First, the inference difficulty is intrinsic to the data at each GD, which cannot be known prior to task execution. Although the resulting decision-making problem can be cast as a partially observable Markov decision process (POMDP)~\cite{sutton1998reinforcement}, conventional approaches typically rely on trial-and-error, e.g., standard deep reinforcement learning (DRL) algorithms, to approximate the optimal policy. While such methods can capture the long-term data statistics, they are unable to exploit \textit{sample-wise difficulty} at runtime, leading to potential under- or over-provisioning of system resources. Second, UAV trajectory control and resource allocation for AI inference are operating on different timescales, giving rise to a \textit{multi-timescale optimization} problem. Specifically, the UAV trajectory must be adjusted continuously, whereas cooperative inference spans longer durations. Bridging these two timescales is particularly challenging because slow-timescale offloading plans depend on future channel conditions that are shaped by fast-timescale trajectory decisions yet to be executed. Third, the joint optimization of offloading decisions and feature transmission leads to a combinatorial action space that grows exponentially with the number of GDs, calling for a \textit{scalable algorithm design} that remains computationally tractable in systems with reasonable scales.

\begin{table}[t]
\centering
\caption{Comparison With Existing Works}
\label{tab:comparison}
\renewcommand{\arraystretch}{1.3}
\begin{tabular}{|c|c|c|c|}
\hline
\textbf{Reference} & \makecell{\textbf{Sample-wise} \\ \textbf{Difficulty}} & \makecell{\textbf{Multi-Timescale} \\ \textbf{Optimization}} & \makecell{\textbf{Scalable} \\ \textbf{Algorithm Design}} \\
\hline
\cite{dingzhu2024multi,jiaocheng2025energy}       & $\times$ & \checkmark & $\times$ \\
\hline
\cite{lixian2024optimal,jingfeng2025dynamic}     & $\times$ & $\times$   & $\times$ \\
\hline
\cite{zhou2025communication} & \checkmark & $\times$ & $\times$ \\
\hline
\cite{yangfu2024ai}          & $\times$ & \checkmark & $\times$ \\
\hline
\textbf{This work}           & \checkmark & \checkmark & \checkmark \\
\hline
\end{tabular}
\end{table}

\subsection{Contributions}
In this paper, we propose a novel UAV-assisted cooperative edge inference framework tailored for LAE scenarios, where the UAV performs pre-assigned LAE duties while simultaneously assisting GDs with their AI inference tasks. Extending our work in~\cite{zhuang2025gc} that considers a static base station, we employ a UAV as a mobile edge server, resulting in the multi-timescale optimization of UAV trajectory control and inference resource allocation. Building on this, we jointly optimize the UAV trajectory, offloading decisions and feature compression ratios to maximize inference accuracy while satisfying the LAE duty requirements. The main contributions of this paper are summarized as follows:
\begin{itemize}
    \item We introduce a trajectory deviation threshold to delineate the acceptable range of divergence from predefined flight paths of the primary LAE duties. Based on this metric, we formulate a joint optimization problem to maximize the averaged inference accuracy of GDs while guaranteeing trajectory deviation remains below the LAE-mandated threshold. To perform online decision-making under varying inference difficulty of individual data and wireless channel condition, the problem is modeled as a constrained POMDP with tractable measurements of inference difficulty at individual GDs as observations.
    \item We design a hierarchical DRL framework to disentangle the slow inference decisions from trajectory optimization in a faster timescale. By leveraging a unified critic network to evaluate actions across different temporal granularities, this framework ensures coherent optimization of UAV mobility and inference resource allocation.
    \item We further propose a novel policy network architecture based on the mixture-of-experts (MoE) paradigm that factorizes the high-dimensional joint action space of inference task offloading and feature compression decisions into specialized gating and expert networks. This decomposition reduces the action space complexity from exponential to linear relative to the number of GDs, thereby enhancing the learning efficiency.
    \item Through extensive simulations, we demonstrate that the proposed algorithm outperforms the baselines in inference accuracy while adhering to the trajectory deviation constraint, achieving an up to 2.96\% improvement in the successful offloading ratio. Furthermore, the proposed algorithm exhibits superior scalability to scenarios with larger numbers of GDs and requires fewer training episodes to converge, validating the learning efficiency of the proposed MoE-based policy network architecture.
\end{itemize}

The remainder of this paper is organized as follows. Section~II describes the system model for the UAV-assisted cooperative edge inference framework. In Section~III, the joint optimization problem is formulated as a constrained POMDP. Section~IV presents the proposed HDRL-MoE framework for solving the formulated problem. Simulation results and analysis are provided in Section~V, followed by conclusions in Section~VI.

\section{System Model}

\begin{figure*}
    \centering
    \includegraphics[width=0.70\textwidth]{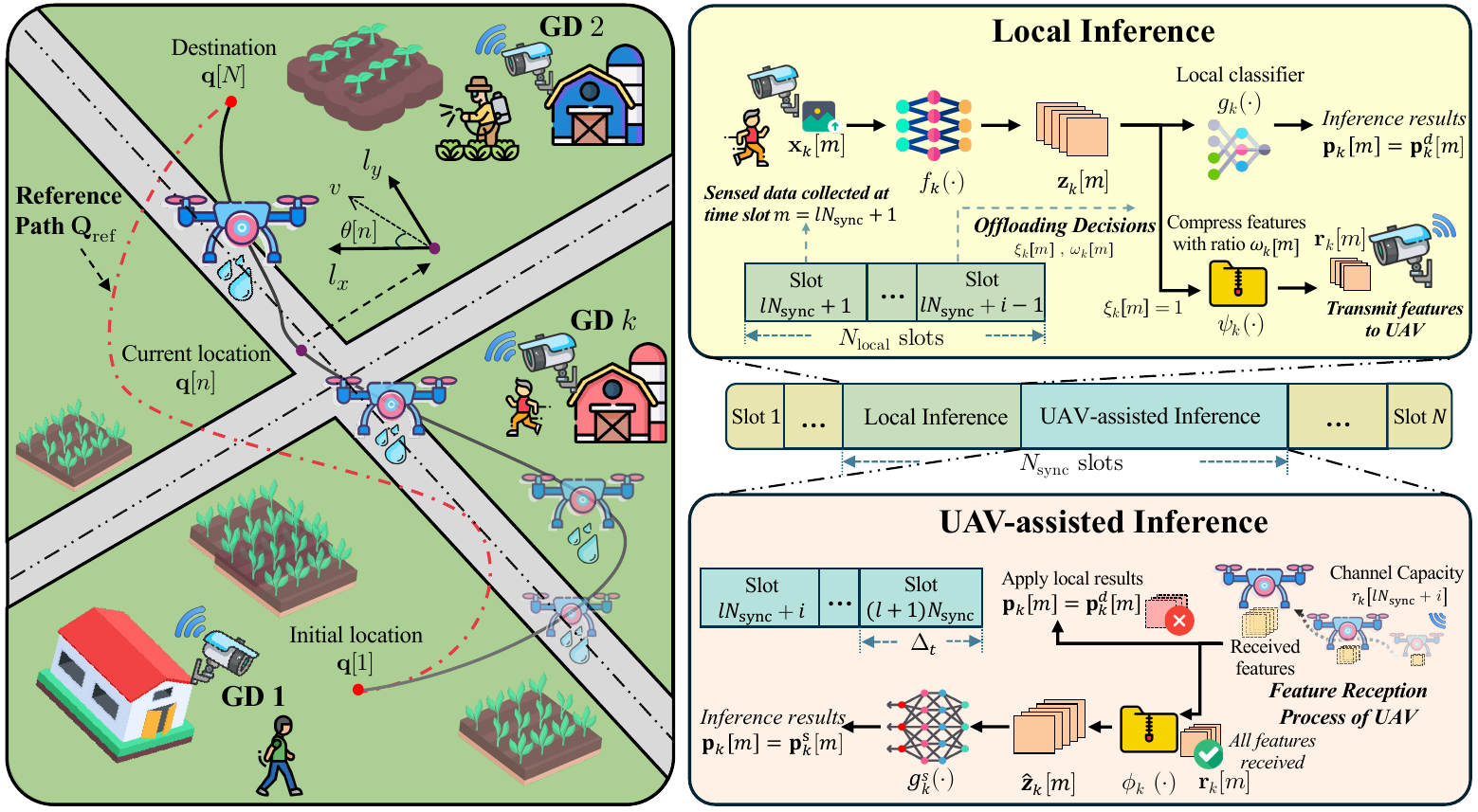}
    \caption{System model of UAV-assisted cooperative edge inference for LAE.}
    \label{fig:system_model}
    \vspace{-1em}
\end{figure*}

We consider a UAV-assisted intelligent wireless network for LAE as depicted in Fig.~\ref{fig:system_model}, where a UAV performs pre-assigned LAE duties, such as surveillance and patrolling, over mission cycles of duration $T$. 
{There are $K$ GDs in the network that need to execute independent inference tasks, e.g., crop classification in precision agriculture and object detection in smart cities.} Note that while the UAV is primarily responsible for LAE services, it also serves as an aerial edge server~\cite{jingfeng2025dynamic} to assist the GDs with their inference tasks.
In this paper, we focus on a mission cycle that is divided into $N$ discrete time slots. Denote $\mathcal{K} \triangleq \{1, 2, \ldots, K\}$ and $\mathcal{N} \triangleq \{1, 2, \ldots, N\}$ as the set of GDs and discrete time slots in a mission cycle, respectively.
{The GDs are static and the two-dimensional coordinates of GD $k$ is given as $[l_{x, k}, l_{y, k}]$. }

\subsection{UAV Flight Model}
We assume the UAV cruises at a fixed altitude, denoted as $H$, with constant speed $v$. The horizontal flight direction of the UAV at time slot $n$ is determined by its heading angle $\theta[n] \in [-\pi, \pi]$. 
{Denote the horizontal coordinates of the UAV at time slot $n$ as $\mathbf{q}[n] \triangleq [l_x[n], l_y[n]]^{T} \in \mathbb{R}^2$, which evolves as follows:}
\begin{align}
    \mathbf{q}[n+1] &= \mathbf{q}[n] + v \cdot \begin{bmatrix}
        \cos \theta[n] \\
        \sin \theta[n]
    \end{bmatrix} \cdot \Delta_t, n = 1, \cdots, N-1,
\end{align}
where $\Delta_t \triangleq T \slash N$ denotes the duration of each time slot. Let $\mathbf{Q} \triangleq [\mathbf{q}[1],\cdots,\mathbf{q}[N]] \in \mathbb{R}^{2\times N}$ be the UAV's positions over the $N$ time slots. It is also assumed that the UAV departs from the initial location $\mathbf{q}[1] = [l_{x, \text{init}}, l_{y, \text{init}}]^{T}$ at the start of the mission cycle and it should arrive at the destination $\mathbf{q}[N] = [l_{x, \text{end}}, l_{y, \text{end}}]^{T}$ at the end of the cycle. The UAV's location during each time slot can be regarded as unchanged by choosing $N$ to be large enough, since the UAV movement within a sufficiently short time interval is negligible.

A reference path given by $\mathbf{Q}_{\text{ref}} \triangleq [\mathbf{q}_{\text{ref}}[1], \ldots, \mathbf{q}_{\text{ref}}[N]]$ is pre-determined according to the LAE duties, where $\mathbf{q}_{\text{ref}}[n]$ denotes the reference position of the UAV at time slot $n$. For instance, the reference path of a patrolling UAV should ensure that critical checkpoints are traversed at regular intervals~\cite{cuong2022uav, deng2023beamforming}. However, such a predefined trajectory may not be efficient for the collaborations between the UAV and GDs. Specifically, severe signal attenuations caused by long propagation distances may prevent a GD located far away from the predefined path from offloading its inference tasks to the UAV. As such, we ease the requirement of strictly following the predefined trajectory and allow the UAV flight along an optimized path that balances the interests of LAE duties and inference tasks. 

To safeguard the UAV being compatible with LAE duties, we measure the deviation of the UAV trajectory from the reference path as follows:
\begin{align}
d_{\text{dev}}(\mathbf{Q}) \triangleq \sqrt{\frac{1}{N}\sum_{n=1}^{N}\left\|\mathbf{q}[n]-\mathbf{q}_{\text{ref}}[n]\right\|^2 }.
\end{align}
where $\left\| \cdot \right\|$ denotes the Euclidean norm. 
The path deviation metric $d_{\text{dev}}(\cdot)$ plays a pivotal role in our system, as excessive detours could disrupt the normal operations of primary LAE duties. 
Take crop spraying as an example. Minor deviations from the reference path are tolerable because droplet dispersion and planned spray overlap ensure the UAV can still cover the entire target area~\cite{Viridiana2022spray}. However, significant deviations could lead to missed coverage and diminish the effectiveness of the mission.
Similarly, in low-altitude surveillance, slight path variations may even help enhance the level of intelligence as richer features can be captured, but substantial deviations could result in untreated areas~\cite{fang2024strategies}.

\subsection{Inference Task Model}
We assume that the GDs sense the environments with embedded sensors (such as cameras) every $N_{\text{sync}}$ time slots. They then infer the class of sensed data to obtain a timely understanding on their surroundings prior to the next sensing time slot. For simplicity, the GDs are assumed to act synchronously. Specifically, they sense at time slot $ m \triangleq l N_{\text{sync}}+1, l = 0, 1, \cdots, \lfloor (N-1) / N_{\text{sync}} \rfloor$. The sensed data of GD $k$ is denoted as $\mathbf{x}_{k}[m]\in\mathbb{R}^{d^{\text{in}}_{k}}$ with $d_{k}^{\text{in}}$ representing its dimension. We denote the set of time slots where the GDs sense as $\mathcal{M}$, which has cardinality $\lfloor (N-1) / N_{\text{sync}} \rfloor+1$. The set of possible classes at GD $k$ is denoted as $\mathcal{C}_{k}$.

\subsection{UAV-Assisted Cooperative Inference}
\label{sec.coinference}
To predict the classes of sensed data samples, each GD may deploy a pre-trained deep neural network (DNN) model. However, due to the limited on-device computational capacities, only lightweight DNN models can be supported, which bottlenecks the inference accuracy. This prompts the aid of the UAV, on which, more powerful DNN models can be deployed. Unfortunately, some data samples are sizable and directly offloading them to the UAV shall incur heavy communication overhead~\cite{zhou2025communication}. In addition, as a movable platform, the UAV is also resource-constrained, e.g., with limited storage space and processing power, which prohibits it from accommodating many state-of-the-art (SOTA) DNN models simultaneously.

To address these challenges, we turn to the cooperative edge inference framework~\cite{shao2023task}, which splits a SOTA DNN model into two partitions, with one partition deployed on the GD and the other on the UAV. The on-device model partition, denoted as $f_k(\cdot)$, extracts $d_{k}$-dimensional intermediate features from local data as follows:
\begin{align}
	\mathbf{z}_{k}[m] = f_{k}(\mathbf{x}_{k}[m]), m\in\mathcal{M},
	\label{local_feature}
\end{align}
of which each dimension is encoded by $e_k$ bits. In contrast to the processing pipelines of most cooperative edge inference frameworks that always transmit the intermediate features for remote analysis, we introduce a local classifier, denoted as $g^{d}_{k}(\cdot)$, at GD $k$ to obtain preliminary classification results, which are given in terms of logits, i.e., the probabilities of a data sample belonging to different classes, as follows:
\begin{align}
\mathbf{p}^{d}_{k}[m] = \text{Softmax}\left(g^{d}_{k}(\mathbf{z}_k[m])\right) \in [0,1]^{|\mathcal{C}_{k}|}, m\in\mathcal{M}.
\end{align}
Our design is inspired by the early-exit DNN architectures~\cite{rahmath2024early,rongkang2022resource} that can bring dual benefits for UAV-assisted cooperative AI inference. On one hand, when the UAV's aid is not available, a GD can still obtain classification results despite with possibly lower accuracy. On the other hand, with $\mathbf{p}^{d}_{k}[m]$, one may infer the classification difficulty of a data sample that helps determine whether the intermediate features should be further processed by the UAV-based model partition, denoted as 
$g^{s}_{k}(\cdot)$. 
The offloading decision, denoted as $\xi_{k}[m]\in\{0,1\}$, is made for each data sample, where $\xi_{k}[m] = 1$ indicates that GD $k$ is selected for offloading the intermediate feature, and $\xi_{k}[m] = 0$ implies that GD $k$ will rely solely on its local classifier. When $\xi_{k}[m] = 0$, the final logits of the classification result for GD $k$ is determined as $\mathbf{p}_{k}[m]=\mathbf{p}^{d}_{k}[m]$. 

We assume that the offloading decision is made after every GD has produced its preliminary classification results for the current synchronization period. Specifically, the first $N_{\text{local}} < N_{\text{sync}}$ time slots of each period are reserved for local inference at the GDs, ensuring that the decision-making process is based on complete and up-to-date local prediction outcomes. This timing is critical as it allows the system to assess the relative classification difficulty across GDs according to $\mathbf{p}^{d}_{k}[m]$. Consequently, samples exhibiting lower confidence are prioritized for offloading to the UAV by leveraging the superior predictive power of the UAV-based model. This difficulty-aware strategy ensures channel resources are allocated preferentially to samples where UAV processing yields the highest marginal accuracy gain, thereby optimizing spectral efficiency. We note that such an offloading strategy is particularly effective in systems where the GDs exhibit similar local inference latency, for instance, when they perform similar classification tasks and possess comparable computational capabilities. A representative example is multi-camera multi-object tracking deployments~\cite{nguyen2022multi}, in which cameras monitor different parts of the environment while running the same model variant. 
In general scenarios, $N_{\text{sync}}$ should be determined according to the slowest local inference latency among all GDs to ensure that the offloading decisions are made based on the most recent local inference results from all GDs.

Moreover, if GD $k$ is selected to be assisted by the UAV, the intermediate feature $\mathbf{z}_k[m]$ is first compressed by removing redundant dimensions with the least magnitudes~\cite{molchanov2017pruning}. The compressed intermediate feature is expressed as follows:
\begin{align}
	\mathbf{r}_k[m] = \psi_{k} (\mathbf{z}_k[m], \omega_k[m]), m \in \mathcal{M},
\end{align}
where $\psi_{k}(\cdot)$ denotes the compressor for GD $k$ and $\omega_k[m] \in [0, 1]$ represents the compression ratio. 
Note that a smaller value of $\omega_{k}[m]$ leads to more compressed features and the feasible set of $\omega_k[m]$ is given as $\bm{\Omega}_k \triangleq \{0, \Omega_{k}[1],\cdots,\Omega_{k}[|\bm{\Omega}_{k}|-1]\}$. Specifically, $\omega_{k}[m]=0$ corresponds to the case where GD $k$ forgoes offloading its intermediate features even if it is selected for UAV assistance, which typically occurs when channel conditions are poor or local inference is sufficient. In this case, GD $k$ falls back to the local inference result, i.e., $\mathbf{p}_{k}[m] = \mathbf{p}^{d}_{k}[m]$. Otherwise, the compressed intermediate feature $\mathbf{r}_k [m]$, comprising $b_k[m] \triangleq \omega_k[m] d_{k} e_{k}$ bits, is transmitted to the UAV during the remaining time slots of the synchronization period.

Upon successful reception of the intermediate feature from GD $k$, i.e., $\sum_{i=N_{\text{local}}}^{N_{\text{sync}} - 1} r_{k}[m + i] \Delta_t \geq b_k[m]$, the UAV perform upsampling via zero interpolation to recover the original intermediate feature $\mathbf{z}_k[m]$ before feeding it into the UAV-based model, which is expressed as:
\begin{align}
	\hat{\mathbf{z}}_k[m] = \phi_{k}(\mathbf{r}_k[m], \omega_k[m])
\end{align}
with $\phi_{k}(\cdot)$ denoting the upsampling operation at GD $k$. The UAV-based model then maps the input $\hat{\mathbf{z}}_k[m]$ to the final logits $\mathbf{p}_{k}[m] = \mathbf{p}^{s}_{k}[m] = \text{Softmax}\left(g^{s}_{k}(\hat{\mathbf{z}}_k[m])\right) \in [0,1]^{|C_{k}|}$. 
In our design, we assume that the latency of UAV-based inference is negligible compared with the communication latency, which is reasonable when the UAV is equipped with sufficiently powerful onboard hardware. Nevertheless, the framework can be readily extended to account for non-negligible UAV-side inference latency by reserving several time slots at the end of each synchronization period for the UAV to complete its processing tasks.
Finally, the class of $\mathbf{x}_{k}[m]$ is determined as $\hat{y}_{k}[m] = \mathop{\arg \max}_{c\in\{1,2,\cdots,C_{k}\}} \left\{\mathbf{p}_{k,c}[m]\right\}$ with $\mathbf{p}_{k,c}[m]$ denoting the $c$-th dimension of $\mathbf{p}_{k}[m]$. 
Due to the more abundant resources available on UAV, a more capable inference model $g^{s}_{k}(\cdot)$, such as one with a deeper network architecture, can be deployed compared to the on-device model $g^{d}_{k}(\cdot)$ at GD $k$, thereby achieving higher classification accuracy through cooperative inference.

\subsection{Uplink Channel Model}
Let $d_{k}[n] \triangleq \sqrt{(l_x[n] - l_{x, k})^2 + (l_y[n] - l_{y, k})^2 + H^2}$ denote the distance between the UAV and GD $k$ at time slot $n$. Assuming each GD is allocated an orthogonal channel for uplink transmission following a free-space path loss model, the uplink signal-to-noise ratio (SNR) from GD $k$ to the UAV at time slot $n$ is given by
\begin{align}
    \gamma_{k}[n] = \frac{\lambda_k p_k}{\sigma^2 d_{k}[n]^{2}}, k \in \mathcal{K}, n \in \mathcal{N},
\end{align}
where $p_k$ and $\lambda_k$ denote the transmit power and channel coefficient of GD $k$ at the reference distance $d_0 = 1$ m, respectively, and $\sigma^2$ is the additive Gaussian noise power. The achievable uplink data rate from GD $k$ to the UAV at time slot $n$ can be expressed as follows:
\begin{align}
    r_{k}[n] = B \log_2(1 + \gamma_{k}[n]),
\end{align}
where $B$ is the channel bandwidth allocated to each GD.
In light of the constrained computational resources at the UAV, a limitation is imposed such that at most $U$ GDs can simultaneously offload their inference tasks, i.e., $\sum_{k=1}^{K} \xi_{k}[m] \leq U$. 
With the intermediate features, the UAV executes its role in the cooperative inference scheme by applying the UAV-based model $g^{s}_k(\cdot)$ to generate refined logits at the end of each synchronization period to determine the class of $\mathbf{x}_{k}\left[m\right]$, as described in Sec.~\ref{sec.coinference}. 

\section{Problem Formulation}
\label{sec:problem_formulation}
In this section, we formulate a joint optimization problem that coordinates the UAV trajectory, the offloading decisions of GDs, and their respective compression ratios. To account for the dynamic nature of the environment, we cast the problem under the framework of Markov decision process (MDP). Within this framework, the UAV makes sequential decisions in adaptation to the temporal variations in the sensed data for optimized system performance.

The objective of the optimization problem is to maximize the classification accuracy across all GDs while minimizing the UAV trajectory deviation from the predefined reference path. This optimization problem presents several key challenges.
First, the UAV trajectory is tightly coupled with the task offloading and feature transmission policies. Specifically, to ensure successful intermediate feature transmission, the UAV must adjust its flight path to maintain favorable channel conditions with the GDs. However, the offloading decisions are made at a slower timescale compared to the UAV trajectory control, leading to a timescale mismatch that complicates their joint optimization.
Second, the relationship between the offloading decisions and classification accuracy is highly complex and cannot be analytically characterized, which arises from the heterogeneous sensitivity of inference tasks to feature fidelity.
Finally, the UAV trajectory should inherently be determined sequentially over time, and the temporal characteristics of the sensed data across GDs are unknown a priori. 

To formulate the MDP problem, we make the assumption below regarding the statistical properties of the sensed data.
\begin{assumption}
	{The inference data $\{\mathbf{x}_k[m]\}$ and ground-truth labels $\{y_k[m]\}$ for GD $k$ are independently sampled from the joint distribution $P_{k}(\mathbf{X}_k, Y_k)$.}
	\label{assumption:1}
\end{assumption}
Since the proposed UAV-assisted cooperative edge inference system operates over a finite number of $N$ time slots, the underlying decision-making process naturally follows an episodic structure~\cite{sutton1998reinforcement}. In addition, the constraints on the maximum number of GDs that can offload simultaneously, along with the hard requirements on the initial and destination locations of the UAV, restrict the feasible action space. Therefore, the problem is characterized as a partial observable \textit{episodic constrained MDP}, which can be represented by a five-tuple $\mathcal{M} = (\mathcal{S}, \mathcal{O}, \mathcal{A}, \mathcal{R}, \mathcal{P})$ defined as follows:
\begin{enumerate}
    \item \textbf{State Space $\mathcal{S}$:} In an MDP, the state should encapsulate all relevant system information required for decision-making. Thus, the system state at time slot $n$ is defined as $\mathbf{s}[n] \triangleq \left\{n, \mathbf{q}[n], \mathbf{X}[m^{\prime}], \mathbf{Y}[m^{\prime}], \mathbf{P}[m^{\prime}] \right\}$ with $m^{\prime} \triangleq N_{\text{sync}} \left\lfloor {(n-1)} \slash {N_{\text{sync}}} \right\rfloor + 1$ denoting the index of the most recent sensing time slot at GDs. 
    Moreover, $\mathbf{X}[m^{\prime}] \triangleq \left\{\mathbf{x}_{1}[m^{\prime}], \ldots, \mathbf{x}_{K}[m^{\prime}]\right\}$, $\mathbf{Y}[m^{\prime}] \triangleq \left\{y_{1}[m^{\prime}], \ldots, y_{K}[m^{\prime}]\right\}$, and $\mathbf{P}[m^{\prime}] \triangleq \left\{\mathbf{p}_{1}[m^{\prime}], \ldots, \mathbf{p}_{K}[m^{\prime}]\right\}$ denote the sensed data of all GDs, the corresponding ground truth labels, and the classification results, respectively. 
    In other words, we are only concerned with the latest sensed data at the GDs at each time slot due to \textbf{Assumption~\ref{assumption:1}}.
    \item \textbf{Observation Space $\mathcal{O}$:} In the considered system, the UAV does not have access to the actual sensed data $\mathbf{X}[m^{\prime}]$ and ground-truth labels $\mathbf{Y}[m^{\prime}]$ at the GDs. However, information regarding the classification difficulty of the data is essential for enhancing inference performance. To this end, we employ the entropy of logits obtained by the local model at the GDs, i.e., $t_{k}[m^{\prime}] = - \sum_{c=1}^{C_{k}} \mathbf{p}^{d}_{k,c}[m^{\prime}] \log (\mathbf{p}^{d}_{k,c}[m^{\prime}]), k \in \mathcal{K}$, to represent the classification difficulty~\cite{settles2009active}. 
    Thus, the observation at time slot $n$ is defined as $\mathbf{o}[n] \triangleq \left\{n, \mathbf{q}[n], \mathbf{t}[m^{\prime}] \right\}$ with $\mathbf{t}[m^{\prime}] \triangleq \left\{t_{1}[m^{\prime}], \ldots, t_{K}[m^{\prime}]\right\}$.

    \item \textbf{Action Space $\mathcal{A}$:} The action at time slot $n$ is defined as $\mathbf{a}[n] \triangleq \left\{\theta[n], \bm{\xi}[m^{\prime}], \bm{\omega}[m^{\prime}]\right\}$, where $\bm{\xi}[m^{\prime}]$ and $\bm{\omega}[m^{\prime}]$ are the offloading indicators of all GDs and their corresponding compression ratio decisions, respectively. 
    Due to the timescale mismatch between trajectory control and offloading decisions, the three components of $\mathbf{a}[n]$ are updated at different frequencies. 
    \begin{itemize}
        \item The offloading indicators $\bm{\xi}[m^{\prime}]$ and compression ratios $\bm{\omega}[m^{\prime}]$ are only updated once per synchronization period. First, during the on-device processing, i.e., $n = m^{\prime}, \ldots, m^{\prime} + N_{\text{local}} - 1$, both the offloading indicators and compression ratios are set to zero. Instead, they are updated at the beginning of the UAV processing, i.e., at time slot $n = m^{\prime} + N_{\text{local}}$, and remain unchanged for the rest of the synchronization period, i.e., $n = m^{\prime} + N_{\text{local}}, \ldots, m^{\prime}+N_{\text{sync}} - 1$.
        \item The UAV heading angle $\theta[n]$ is updated at every time slot $n$ to balance between the performance of primary LAE duties and the inference tasks.
    \end{itemize}
    The action space is defined as $\mathcal{A} = \mathcal{A}_{\theta} \times \mathcal{A}_{\xi} \times \mathcal{A}_{\omega}$, where $\mathcal{A}_{\theta} = [-\pi, \pi]$ denotes the possible heading angles, and $\mathcal{A}_{\xi} = \{0, 1\}^{K}$ indicates that the action space for the offloading indicators take Boolean values. 
    Notably, not all combinations of the UAV heading angle, offloading indicators, and compression ratios are feasible due to the system constraints. Specifically, the number of simultaneously offloading GDs should not exceed a predefined limit, i.e., $\sum_{k=1}^{K} \xi_{k}[m^{\prime}] \leq U$, and the initial and final locations of the UAV need to be satisfied.

    \item \textbf{Reward Function:} The reward function is defined as:
    \begin{align}
        r[n] &= - \sum_{k=1}^{K} \mathcal{L}_a(\mathbf{p}_{k}[m^{\prime}], {y}_{k}[m^{\prime}]) \delta(n - m^{\prime} - N_{\text{sync}} + 1) \nonumber \\
        & \quad - \lambda  \mathcal{L}_d(\mathbf{q}[n]), 
    \end{align}
    where the Kronecker delta function $\delta(\cdot)$ ensures that the inference performance is accounted at the end of each synchronization period. Moreover, $\lambda \geq 0$ is a weighting factor that balances the classification accuracy and trajectory deviation. 
    In addition, $ \mathcal{L}_d(\mathbf{q}[n]) \triangleq \left\| \mathbf{q}[n] - \mathbf{q}_{\text{ref}}[n] \right\|^2$ measures the deviation of the UAV from the reference checkpoint at time slot $n$,
    and $\mathcal{L}_a(\mathbf{p}_{k}[m^{\prime}], {y}_{k}[m^{\prime}]) \triangleq -\sum_{c=1}^{C_{k}} y_{k,c}[m']\log(\mathbf{p}_{k,c}[m'])$ is the cross entropy loss function, which measures the classification accuracy of the UAV-assisted inference on the sensed data collected at time slot $m$. 
    Unlike binary classification metrics, the cross-entropy loss captures predictive confidence in addition to correctness~\cite{farebrother2024stop}. Therefore, challenging samples that yield low-confidence predictions may still incur high penalties even if they are correctly classified.
    By incorporating cross-entropy into the reward design, the UAV is incentivized to prioritize data samples that are more challenging to classify, thereby improving the reliability and accuracy of the inference system.

    \item \textbf{State Transition Probability:} The state transition probability is denoted as $\mathcal{P}(\mathbf{s}[n+1] | \mathbf{s}[n], \mathbf{a}[n])$, which captures the chance of having the state $\mathbf{s}[n+1]$ at time slot $n+1$ given the current state $\mathbf{s}[n]$ and action $\mathbf{a}[n]$. Specifically, the transition can be factorized as
    \begin{align}
        & \mathcal{P}(\mathbf{s}[n+1] | \mathbf{s}[n], \mathbf{a}[n]) =   \mathcal{P}(\mathbf{q}[n+1] | \mathbf{s}[n], \mathbf{a}[n])  \nonumber \\
        & \quad \cdot \mathcal{P}({\mathbf{X}[m^{\prime\prime}], \mathbf{Y}[m^{\prime \prime}], \mathbf{P}[m^{\prime \prime}] | \mathbf{s}[n], \mathbf{a}[n]}),
    \end{align}
    where $m^{\prime\prime} = N_{\text{sync}} \left\lfloor {n} \slash {N_{\text{sync}}} \right\rfloor + 1$ denotes the index of the most recent sensing time slot before time slot $n+1$. 
    The first term on the right-hand side represents the deterministic outcome of the UAV location and uplink transmission rates given the current state and action, while the second term denotes the conditional distribution of the sensed data. 
    However, because the distributional information of the sensed data is unavailable, the relationship among $\mathbf{P}[m^{\prime \prime}]$, $\mathbf{s}[n]$, and $\mathbf{a}[n]$ is hard to model, which renders the closed-form expression of the transition intractable.
\end{enumerate}

The objective of the MDP formulation is to find a policy $\pi: \mathcal{O} \rightarrow \mathcal{A}$ that maximizes the expected cumulative reward over the mission time $T$. 
However, solving the problem is non-trivial due to the following reasons. 
First, as the statistical properties of sensed data are unknown a priori, the state becomes partially observable. 
Second, the action space encompasses both continuous variables and discrete variables. This hybrid continuous-discrete space further complicates the optimization process as conventional MDP solvers, e.g., reinforcement learning algorithms, are typically tailored to either continuous or discrete action spaces. 
Moreover, the mismatch in timescales between UAV trajectory control and inference decisions exacerbates the challenge. 
Third, the state and action spaces are high-dimensional and scales with the number of GDs $K$. 
Therefore, an efficient and scalable algorithm is imperative to manage the complexity that scales with both the number of GDs $K$.
For these reasons, we develop a novel DRL-based approach in the next section.

\section{Proposed HDRL-MoE Framework}
To solve the formulated constrained episodic MDP problem in Sec.~\ref{sec:problem_formulation}, we propose \textbf{HDRL-MoE}, a DRL framework for joint optimization of UAV trajectory, task offloading, and feature transmission policies. Our approach addresses the identified key challenges. Specifically, the DRL framework in \textbf{HDRL-MoE} is designed to tackle the partial observability of the environment by leveraging an \textit{experience replay mechanism} that captures the state transitions.
In addition, we handle the hybrid continuous-discrete action space and timescale mismatch issue through a \textit{hierarchical architecture} that optimizes UAV trajectory and collaborative inference policies at distinct temporal scales. This approach utilizes separate actor networks for continuous UAV trajectory control and discrete inference decisions, allowing each to specialize in its respective task. A unified critic network is employed to coordinate the learning process of the two sets of interdepedents. To resolve the scalability challenge in low-altitude wireless networks with many GDs, we further propose a novel \textit{MoE-based architecture}. This architecture employs a lightweight network to determine offloading decisions for all GDs, while assigning a dedicated expert network to each selected GD to optimize its compression ratio in task offloading. The high-dimensional decision-making problem is thus decomposed into multiple low-dimensional subproblems that can be handled with smaller networks.
\begin{figure*}
    \centering
    \includegraphics[width=0.75\textwidth]{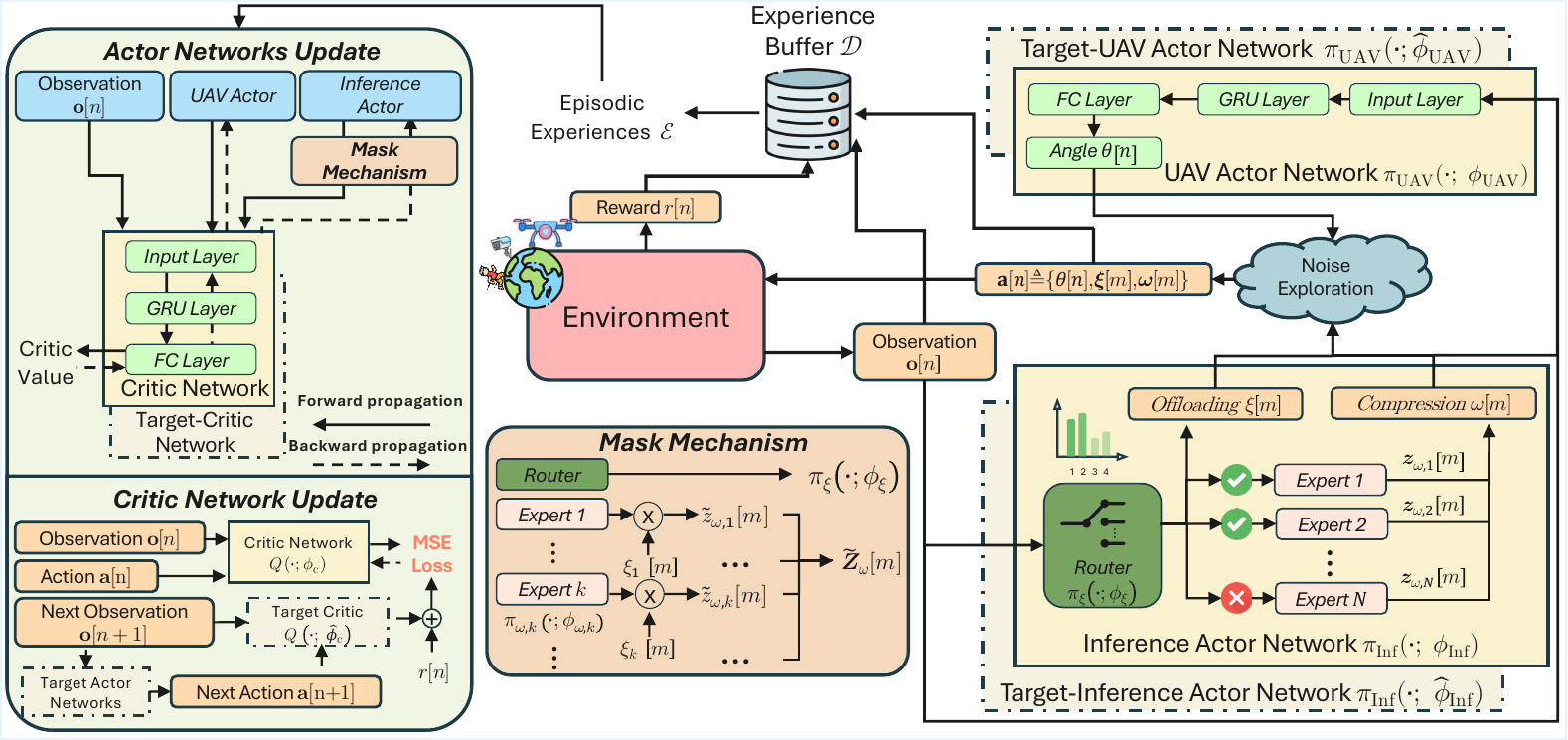}
    \caption{Overview of the proposed HDRL-MoE framework.}
    \label{fig:algorithm}
    \vspace{-1em}
\end{figure*}

\subsection{Hierarchical DRL Framework}
As depicted in Fig.~\ref{fig:algorithm}, the proposed framework employs a hierarchical DRL architecture to effectively manage the inherent multi-timescale decision-making process in UAV-assisted cooperative edge inference systems, where the offloading indicators $\bm{\xi}[m]$ and compression ratios $\bm{\omega}[m]$ are determined per synchronization period, while the UAV heading angle $\theta[n]$ is determined at each time slot $n$.
This temporal entanglement necessitates a hierarchical DRL architecture comprising three network modules, namely:
\begin{itemize}
    \item \textbf{Inference Actor Network} $\pi_{\text{Inf}}(\cdot)$: This network periodically configures $\bm{\xi}[m]$ and $\bm{\omega}[m]$ at time slot $m \in \mathcal{M}$. 
    Denote $\phi_{\text{Inf}}$ as the trainable parameters of $\pi_{\text{Inf}}(\cdot)$, the inference actor network operates on the observation $\mathbf{o}[m]$ to output the offloading indicators and compression ratios as follows:
    \begin{align}
        \left[\bm{\xi}[m], \bm{\omega}[m]\right] = \pi_{\text{Inf}}(\mathbf{o}[m]; \phi_{\text{Inf}}).
    \end{align}
    \item \textbf{UAV Actor Network} $\pi_{\text{UAV}}(\cdot)$: This network determines the UAV heading angle $\theta[n]$ at every time slot $n \in \mathcal{N}$. 
    Denote $\phi_{\text{UAV}}$ as the trainable parameters of $\pi_{\text{UAV}}(\cdot)$, the UAV actor network outputs the UAV heading angle as follows:
    \begin{align}
        \theta[n] &= \pi_{\text{UAV}}(\mathbf{o}[n]; \phi_{\text{UAV}}).
    \end{align}
    \item \textbf{Unified Critic Network} $Q(\cdot)$: Denote $\phi_{\text{c}}$ as the trainable parameters of $Q(\cdot)$, we employ the critic network $Q(\mathbf{o}[n], \mathbf{a}[n]; \phi_{\text{c}})$ to approximates the state-action value function, i.e., the expected cumulative reward starting from time slot $n$ given the current observation $\mathbf{o}[n]$ and the actions $\mathbf{a}[n]$ taken by both actor networks. 
    Unlike conventional DRL methods that employ separate critic networks for different actor networks~\cite{mei2021intelligent}, this unified architecture explicitly captures the interdependence between UAV trajectory and inference variables. The joint modeling is crucial for the coordinated optimization of both actor networks.
\end{itemize}
Since the actor network may favor actions with positive estimation noise, overestimation errors accumulate in the critic network as current estimates are used to update future estimates, leading to unstable and suboptimal policy. To mitigate this, we implement separate target networks for both the actor and critic components~\cite{ddqn2016}. In particular, we introduce a set of three target networks, i.e., target-critic network, target-UAV actor network, and target-inference actor network, which are parameterized by $\hat{\phi}_{\text{c}}$, $\hat{\phi}_{\text{UAV}}$, and $\hat{\phi}_{\text{Inf}}$, respectively.
These target networks mirror the architectures of their respective actor networks but are updated through delayed parameter synchronization. The specific update mechanism is detailed in Sec.~\ref{sec:model_update}.

\subsection{Action Selection Policy}

\subsubsection{MoE-based Inference Task Offloading Policy}
The inference task policy determines the offloading decisions and compression ratios of GDs, i.e., $\bm{\xi}[m] \in \left\{0, 1\right\}^{K}$ and $\bm{\omega}[m] \in \bm{\Omega}_{1} \times \cdots \times \bm{\Omega}_{K}$, respectively. 
These variables should be chosen from a discrete action space. Consequently, the inference actor network $\pi_{\text{Inf}}(\cdot)$ is typically constructed to output a categorical probability distribution over these discrete actions in traditional DRL frameworks, e.g., DQN and DDPG~\cite{lillicrap2015continuous}. 
However, the dimension of action space grows exponentially with the number of GDs and feasible compression ratios per GD, i.e., $\mathcal{O}(\prod_{k=1}^{K} \left(|\bm{\Omega}_k|\right))$, rendering the training of the inference actor network formidable.

To address this challenge, we propose a novel two-stage policy network architecture inspired by the MoE paradigm~\cite{molchanov2017pruning}.
The core idea of MoE is the selective activation mechanism that dynamically triggers specific expert networks based on input features instead of invoking the full model. 
This characteristic suits well to our scenario, since only a subset of GDs is required to update their compression ratios in each synchronization period. Our approach decomposes the joint action space of offloading decisions and compression ratios. Specifically, the inference actor network $\pi_{\text{Inf}}(\cdot)$ is constructed by two specialized components, namely a gating network $\pi_{\xi}(\cdot)$ and $K$ expert networks $\pi_{\omega,k}(\cdot)$. The gating network determines the offloading decisions $\bm{\xi}[m]$ and the subset of expert networks to be activated. An activated expert network $\pi_{\omega,k}(\cdot)$ selects the feature compression ratio $\omega_k[m]$ for GD $k$. 

Given the observation $\mathbf{o}[m]$, the gating network generates a vector of logits for the offloading decision $\bm{z}_{\xi}[m] \in \mathbb{R}^{K}$. Each element of the vector quantifies the offloading priority of a corresponding device, which is formulated as:
\begin{align}
    \bm{z}_{\xi}[m] &= \pi_{\xi}(\mathbf{o}[m]; \phi_{\xi}), 
\end{align}
where $\phi_{\xi}$ denotes the trainable parameters of $\pi_{\xi}(\cdot)$. The top-$U$ largest logits in $\bm{z}_{\xi}[m]$ are selected to form the offloading decision vector $\bm{\xi}[m]$, which can be expressed as:
\begin{align}
\xi_k[m] = \begin{cases}
1, & z_{\xi,k}[m] \text{ is among the top-}U \text{ values} \\
0, & \text{otherwise}
\end{cases}, \quad k \in \mathcal{K}, \label{eq:offload_decision}
\end{align}
where $z_{\xi,k}[m]$ denotes the $k$-th dimension of $\bm{z}_{\xi}[m]$. 
This selection mechanism ensures that at most $U$ GDs are considered for offloading, which satisfies the constraint imposed by the UAV computational limitation. Specifically, for each selected GD $k$ (i.e., $\xi_k[m] = 1$), the expert network $\pi_{\omega,k}(\cdot)$ computes a logit vector $\bm{z}_{\omega,k}[m]$ over the feasible compression ratio space~$\bm{\Omega}_k$. The optimal decision is then selected as:
\begin{align}
    \bm{z}_{\omega,k}[m] &= \pi_{\omega,k}(\mathbf{o}[n]; \phi_{\omega,k}) \in \mathbb{R}^{|\bm{\Omega}_k|},
\end{align}
where $i=\mathop{\arg \max}_{j} z_{\omega,k, j}[m]$ with $z_{\omega,k, j}[m]$ denoting the $j$-th dimension of $\bm{z}_{\omega,k}[m]$. Therefore, the trainable parameters of the inference actor network can be written as $\phi_{\text{Inf}} \triangleq \left\{\phi_{\xi}, \phi_{\omega,1}, \ldots, \phi_{\omega, K}\right\}$. Notably, this approach reduces the output dimension of the inference actor network from $\mathcal{O}(\prod_{k=1}^{K} \left(|\bm{\Omega}_k|\right))$ to $\mathcal{O}(K \cdot \max_{k \in \mathcal{K}} |\bm{\Omega}_k|)$.

\subsubsection{UAV Trajectory Policy}
The UAV trajectory policy is responsible for determining the UAV heading angle $\theta[n]$ at each time slot $n$. In our \textbf{HDRL-MoE} framework, the UAV trajectory policy acts according to the inference task offloading policy. 
In other words, the output of the UAV actor network $\theta_{\pi}[n]$ utilizes the current observation $\mathbf{o}[n]$ alongside the offloading decisions $\bm{\xi}[m]$ and compression ratios $\bm{\omega}[m]$ determined at the most recent synchronization time slot $m$ (i.e., $m = \lfloor (n-1) / N_{\text{sync}} \rfloor N_{\text{sync}}+1$). This formulation captures how the instantaneous UAV heading angle adapts to the offloading decisions, which is expressed as follows:
\begin{align}
    \theta_{\pi}[n] = \pi_{\text{UAV}}(\mathbf{o}[n], \bm{\xi}[m], \bm{\omega}[m]; \phi_{\text{UAV}}). 
\end{align}
Moreover, to guarantee destination arrival by mission completion, we enforce the straight-flight behavior when the number of remaining time slots, i.e., $N - n - 1$, is smaller than the minimum number of time slots $T_{\min}[n+1] \triangleq \left\lceil \frac{d_{\text{dest}}[n+1]}{v \cdot \Delta_t} \right\rceil$ required to travel the distance from the next position to the destination, where $d_{\text{dest}}[n] = \sqrt{(l_x[n] - l_{x, \text{end}})^2 + (l_y[n] - l_{y, \text{end}})^2}$. In such cases, the UAV heads directly toward the destination~\cite{xiaowen2026isac}. Thus, the UAV heading angle at time slot $n$ is determined as:
\begin{align}
\theta[n] = \begin{cases}
\tan^{-1}\left( \frac{l_{y, \text{end}} - l_y[n]}{l_{x, \text{end}} - l_x[n]} \right), & \text{if } N - n - 1 \leq T_{\min}[n+1] \\
\theta_{\pi}[n], & \text{otherwise}
\end{cases}.
\end{align}

\subsection{Experience Collection}
In the proposed \textbf{HDRL-MoE} framework, the current observation, actions, the next observation, and the system rewards are collected as experience to train the actor-critic networks.  
However, selecting actions with the highest logit value from the gating network for offloading decision and the expert networks for compression ratio selection would yield deterministic behavior that hinders exploration. 
Therefore, it is crucial to incorporate exploration strategies to encourage diverse action selection during the learning process. 
Specifically, the offloading decisions $\bm{\xi}[m]$ and compression ratios $\bm{\omega}[m]$ are sampled from the categorical distribution defined by the actor network output $\bm{z}_{\xi}[m]$ and $\bm{z}_{\omega,k}[m]$, respectively.
Since categorical sampling is non-differentiable and prevents the back-propagation algorithm from updating the actor networks, we adopt the Gumbel-Softmax trick~\cite{jang2016categorical} to obtain differentiable relaxed action vectors that approximate the corresponding one-hot samples. The reparameterized expressions for the offloading and compression actions are given as follows:
\begin{align}
    \hat{z}_{\xi, k}[m] &= \frac{\exp\left(\frac{z_{\xi,k}[m] + g_{\xi,k}[m]}{\tau_{\xi}}\right)}{\sum_{i=1}^{K} \exp\left(\frac{z_{\xi,i}[m] + g_{\xi,i}[m]}{\tau_{\xi}}\right)}, k \in \mathcal{K}, \label{eq:offload_exploration} \\
    \hat{z}_{\omega, k, i}[m] &= \frac{\exp\left(\frac{z_{\omega,k,i}[m] + g_{\omega,k,i}[m]}{\tau_{\omega}}\right)}{\sum_{j=1}^{|\bm{\Omega}_k|} \exp\left(\frac{z_{\omega,k,j}[m] + g_{\omega,k,j}[m]}{\tau_{\omega}}\right)}, \nonumber \\
    & \ \ \ \ \ \ \ \ \ \ \ \ k\in \mathcal{K}, i \in \left\{1, \dots, |\bm{\Omega}_k|\right\} \label{eq:compress_exploration},
\end{align}
where $g_{\xi,k}[m]$ and $g_{\omega,k, i}[m]$ denote the Gumbel random variables sampled from the Gumbel distribution. 
Additionally, $\tau_{\xi}$ and $\tau_{\omega}$ are tunable parameters to control the exploration process: 
the sampling policy becomes increasingly deterministic as $\tau_{\xi} \rightarrow 0$ and $\tau_{\omega} \rightarrow 0$, whereas larger values encourage exploration. 
Given the approximated action probabilities, the offloading decisions $\bm{\xi}[m]$ and compression ratios $\bm{\omega}[m]$ during experience collection are determined by selecting the actions with the highest probabilities according to the procedures defined in~\eqref{eq:offload_decision}. The continuous UAV heading angle $\theta[n]$ is also perturbed by Gaussian noise to encourage exploration, which can be expressed as follows:
\begin{align}
    \tilde{\theta}[n] = \theta_{\pi}[n] + r[n], 
	\label{eq:uav_exploration}
\end{align}
where $r[n] \sim \mathcal{N}(0, \sigma_z^{2})$ denotes the Gaussian noise with zero mean and variance $\sigma_z^{2}$. It is then clipped to the feasible range, i.e., ${\theta}[n] = \max\{-\pi, \min\{\tilde{\theta}_{k}[n], \pi\}\}$. 
Moreover, the exploration noise $\sigma_z^2$ and Gumbel-Softmax temperatures $\tau_{\xi}, \tau_{\omega}$ are each annealed by a decay factor $\kappa \in (0, 1)$ over time from their initial values to ensure sufficient exploration in the early stages and stable convergence in the later stages. Therefore, the exploration noise and temperatures at episode $e$ can be expressed as $\sigma_z^2[e] = \kappa^{e} \sigma_z^2[0]$, $\tau_{\xi}[e] = \kappa^{e} \tau_{\xi}[0]$, and $\tau_{\omega}[e] = \kappa^{e} \tau_{\omega}[0]$.

In the proposed \textbf{HDRL-MoE} framework, the policy is evaluated over the entire mission cycle, i.e., the cumulative reward over $N$ time slots. 
Thus, each piece of experience is collected as a tuple $\bm{\upsilon} \triangleq \left\{\mathbf{o}[n], \mathbf{a}[n], \mathbf{o}[n+1], r[n]\right\}_n$. 
The experience tuples are stored in a replay buffer $\mathcal{D}$ with limited capacity. This buffer operates on a first-in-first-out (FIFO) basis, which ensures that the training data remains relevant to the current policy and helps stabilize the learning process~\cite{mnih2015replay}.

\subsection{Model Update}
\label{sec:model_update}
In each training iteration of the actor and critic networks, a mini-batch of experience tuples, denoted as $\mathcal{E} \triangleq \left\{\bm{e} \right\}$, is sampled from the replay buffer $\mathcal{D}$. Due to the MoE-inspired architecture, an expert network is only activated when the corresponding GD is selected for offloading. Therefore, the proposed two-stage inference task offloading policy necessitates conditional gradient propagation~\cite{mohtashami2022masked}. To this end, we introduce a masked gradient propagation mechanism to ensure that $\pi_{\omega,k}(\cdot)$ is only updated when the corresponding GD $k$ is selected for offloading. Specifically, at time slot $n$, the outputs of the gating network $\bm{\xi}[m]$ dynamically mask the compression ratio logits $\bm{z}_{\omega,k}[m]$ for each GD $k$ as follows:
\begin{align}
    \tilde{\bm{z}}_{\omega,k}[m] = \xi_k[m] \cdot \bm{z}_{\omega,k}[m], m \in \mathcal{M}.
\end{align}
The masked logits of all GDs, denoted as $\tilde{\bm{Z}}_{\omega}[m] \triangleq \left\{\tilde{z}_{\omega,k}[m]\right\}_{k=1}^{K}$, are concatenated with the offloading decision logits $\bm{z}_{\xi}[m]$ and the UAV heading angle $\theta[m]$ to form the input to the critic network, which is expressed as:
\begin{align}
   \tilde{\mathbf{a}}[n] &= \left[\theta[n], \bm{z}_{\xi}[m], \tilde{\bm{Z}}_{\omega}[m]\right] \in \mathbb{R}^{1 + K + \sum_{k=1}^{K} |\bm{\Omega}_k|}.
\end{align}
The actor networks are then updated by maximizing the expected cumulative reward estimated by the unified critic network, which can be expressed as:
\begin{align}
    \phi_{\text{a}} \leftarrow \phi_{\text{a}} + \eta_{\text{a}} \nabla_{\phi} \frac{1}{\left|\mathcal{E} \right|} \sum_{\bm{\upsilon} \in \mathcal{E}} Q(\mathbf{o}[n], \tilde{\mathbf{a}}[n]; \phi_{\text{c}}),
    \label{eq:actor_loss}
\end{align}
where $\eta_{\text{a}}$ is the learning rate for the actor networks. 

The parameters of the critic network, i.e., $\phi_{\text{c}}$, are updated by minimizing the mean squared error (MSE) between the predicted state-action value and the target value, which is computed as follows:
\begin{align}
    \mathcal{L}_{\text{c}}(\phi_{\text{c}}) &= \frac{1}{\left|\mathcal{E} \right|} \sum_{\bm{\upsilon} \in \mathcal{E}} \sum_{n=1}^{N} \left[ y[n] - Q(\mathbf{o}[n], \tilde{\mathbf{a}}[n]; \phi_{\text{c}})\right]^2,
    \label{eq:critic_loss}
\end{align}
where the target value $y[n] = r[n] + Q(\mathbf{o}[n+1], \hat{\mathbf{a}}[n+1]; \hat{\phi}_{\text{c}})$ with
\begin{subequations}
\begin{align}
    \hat{\mathbf{a}}[n+1] &= \left[\hat{\theta}[n+1], \bm{\hat{z}}_{\xi}[m^{\prime \prime}], \hat{\bm{Z}}_{\omega}[m^{\prime \prime}]\right], \\
    \hat{\theta}[n+1] &= \pi_{\text{UAV}}(\mathbf{o}[n+1]; \hat{\phi}_{\text{UAV}}), \\
    \hat{\bm{Z}}_{\omega}[m^{\prime \prime}] &= \left\{\xi_k[m^{\prime \prime}] \cdot \pi_{\omega,k}(\mathbf{o}[n+1]; \hat{\phi}_{\omega,k})\right\}_{k=1}^{K}, \\
    \hat{\bm{z}}_{\xi}[m^{\prime \prime}] &= \pi_{\xi}(\mathbf{o}[n+1]; \hat{\phi}_{\xi}),
\end{align}
\end{subequations}
denoting the actions at the time slot $n+1$ produced by the target actor networks. The parameters of the critic network are updated as $\phi_{\text{c}} \leftarrow \phi_{\text{c}} - \eta_{\text{c}} \nabla_{\phi_{\text{c}}} \mathcal{L}_{\text{c}}(\phi_{\text{c}})$ with $\eta_{\text{c}}$ denoting the learning rate of the critic network. In addition, the target networks are updated via the soft update mechanism as follows:
\begin{subequations}
\begin{align}
    \hat{\phi}_{\text{c}} &\leftarrow \tau_{\text{c}} \phi_{\text{c}} + (1 - \tau_{\text{c}}) \hat{\phi}_{\text{c}}, \label{eq:soft_update_a} \\
    \hat{\phi}_{\text{UAV}} &\leftarrow \tau_{\text{a}} \phi_{\text{UAV}} + (1 - \tau_{\text{a}}) \hat{\phi}_{\text{UAV}}, \label{eq:soft_update_b} \\
    \hat{\phi}_{\text{Inf}} &\leftarrow \tau_{\text{a}} \phi_{\text{Inf}} + (1 - \tau_{\text{a}}) \hat{\phi}_{\text{Inf}}, \label{eq:soft_update_c}
\end{align}
\end{subequations}
where $\tau_{\text{c}}$ and $\tau_{\text{a}}$ are the soft update coefficients for the critic and actor networks, respectively. The pseudocode of the proposed \textbf{HDRL-MoE} framework is summarized in Algorithm~\ref{alg:algorithm}.

\begin{algorithm}[t]
\caption{Pseudocode of the proposed \textbf{HDRL-MoE}.}
\label{alg:algorithm}
\begin{algorithmic}[1]
\STATE \textbf{Input:} Initialize actor networks $\pi_{\text{UAV}}(\cdot)$, $\pi_{\text{Inf}}(\cdot)$, critic network $Q(\cdot)$, target networks, and replay buffer $\mathcal{D}$.
\FOR{each training episode}
    \STATE Reset environment and initialize state $\mathbf{s}[1]$.
    \STATE \textbf{\textit{// Experience Collection}}
    \FOR{$n = 1$ to $N$}
        \STATE {\textit{// Collect Sensing Data}}
        \IF{$n \in \mathcal{M}$} 
            \STATE Obtain latest sensed data $\mathbf{X}[n]$, labels $\mathbf{Y}[n]$.
        \ENDIF
        \STATE {\textit{// Update UAV heading angle with exploration}}
        \STATE Select $\theta[n]$ via \eqref{eq:uav_exploration}.

        \STATE {\textit{// Update offloading actions with exploration}}
        \IF{$n = m + N_{\text{local}}$}
            \STATE Select $\bm{\xi}[m]$ and $\bm{\xi}^{\prime}[m]$ via \eqref{eq:offload_decision} and \eqref{eq:offload_exploration}, respectively.
        \ENDIF
        \STATE Execute actions $\mathbf{a}[n]$ and update environment.
        \STATE Obtain observation $\mathbf{o}[n+1]$, compute reward $r[n]$.
        \STATE Store $(\mathbf{s}[n], \mathbf{a}[n], \mathbf{s}[n+1], r[n])$ in $\mathcal{D}$.
    \ENDFOR
    \STATE \textbf{\textit{// Model Update}}
    \STATE Sample a mini-batch $\mathcal{E}$ from $\mathcal{D}$.
    \STATE Update actor and critic networks via \eqref{eq:actor_loss} and \eqref{eq:critic_loss}, respectively.
    \STATE Soft update target networks via \eqref{eq:soft_update_a}-\eqref{eq:soft_update_c}.
\ENDFOR
\STATE \textbf{Output:} Trained actor networks for joint UAV trajectory and cooperative inference control.
\end{algorithmic}
\end{algorithm}

\section{Simulation Results}
In this section, we evaluate the performance of the proposed \textbf{HDRL-MoE} framework for UAV-assisted cooperative edge inference in low-altitude wireless networks. 

\subsection{Experimental Setup}

\begin{enumerate}
    \item \textit{System Parameters}: We consider a square area of $1 \times 1$ km$^2$ with several GDs randomly distributed within the region. The LAE mission duration is $T = 60$ s, discretized into $N = 48$ time slots of equal duration $\delta = 1.25$ s. The UAV operates at a fixed altitude of $H = 100$ m and maintains a constant speed of $v = 30$ m/s. The reference trajectory is a circular path of radius $r = 150$ m centered at the origin $[0, 300]$ m. The UAV is required to start and end its mission at a designated location, specifically $\mathbf{q}[1] = \mathbf{q}[N] = [0, 0]$ m. The uplink communication bandwidth for each GD is configured as $B_k = 1$ MHz and the noise power is $\sigma^2 = N_0 B = -110$ dBm. Each GD transmits with a fixed power of $p_k = 10$ dBm. 
    The uplink channels are modeled by the free-space path loss model, and the reference channel gain is $\lambda_k = -50$ dB. The data sampling interval at each GD is $N_{\text{sync}} = 4$ time slots, with $N_{\text{local}} = 1$ time slot allocated for local inference.
    \item \textit{Inference Tasks}: We evaluate the proposed framework on image classification tasks over the STL-10 dataset~\cite{coates2011analysis}, which consists of 10 classes of images at $96 \times 96$ resolution. 
    The datasets are divided into four parts for: a) training the classification model; b) training the \textbf{HDRL-MoE} framework; c) validation; and d) testing, containing $5{,}000$, $5{,}000$, $1{,}000$, and $1{,}000$ samples, respectively. The feasible compression ratio set for feature compression is configured as $\{0,0.2,0.4,0.55\}$, and we employ a ResNet-18 architecture as the backbone of our DNN model. The model is partitioned such that the on-device feature extractor $f_k(\cdot)$ comprises the first $4$ residual blocks, while remaining layers are executed at the edge server. The local classifier $g_k^d(\cdot)$ is implemented as a three-layer fully connected network with hidden layer dimension set as $128$.
    \item \textit{Baselines}: To demonstrate the performance of the proposed \textbf{HDRL-MoE} framework, we investigate the following baselines for comparison:
    \begin{itemize}
        \item \textbf{Fixed Trajectory~{(FT)}}: The UAV follows the predefined reference trajectory, while the AI inference-related variables, i.e., offloading decisions $\bm{\xi}[m]$ and compression ratios $\bm{\omega}[m]$, are determined by the proposed MoE-based inference actor networks. 
        \item \textbf{Greedy Inference~{(GI)}}: The UAV trajectory is optimized via the \textbf{HDRL-MoE} framework, whereas inference decisions are made using a greedy heuristic algorithm. Specifically, at each inference decision time slot, i.e., $n \in \mathcal{M}$, the $U$ GDs with the highest local prediction entropy are selected for offloading, and each selected GD adopts the highest compression ratio that can be supported by its current uplink transmission rate. The heuristic approach prioritizes offloading data that are difficult for local inference.
        \item \textbf{Hierarchical DRL~{(HDRL)}}: This scheme shares the same hierarchical DRL framework as \textbf{HDRL-MoE} but replaces the MoE-based inference actor network with a single network that directly outputs all offloading decisions and compression ratios.
        \item \textbf{Uncertainty-Excluded HDRL~{(HDRL-UE)}}: This scheme is an \textbf{HDRL-MoE} variant that excludes the local prediction entropy from the system state in the DRL formulation, disabling  the inference actor network from leveraging the hint of the inference data difficulty.
    \end{itemize}
\end{enumerate}

\subsection{Hyperparameter Settings}

\begin{table}[t]
\caption{Default hyperparameter settings for training the proposed \textbf{HDRL-MoE} framework.}
\label{tab:hyperparameters}
\centering
\renewcommand{\arraystretch}{1.2}
\begin{tabular}{l|c|l|c}
\hline
\textbf{Parameter} & \textbf{Value} & \textbf{Parameter} & \textbf{Value} \\
\hline
Learning rate ($\eta_{\text{a}}$) & 1e-4 & Learning rate ($\eta_{\text{c}}$) & 1e-3 \\
Update coefficient ($\tau_{\text{a}}$) & $0.001$ & Update coefficient ($\tau_{\text{c}}$) & $0.001$ \\
Softmax temp. ($\tau_{\xi}[0]$) & $0.5$ & Softmax temp. ($\tau_{\omega}[0]$) & $0.5$ \\
Batch size ($|\mathcal{E}|$) & $64$ & UAV noise var. ($\sigma_z^2[0]$) & $0.1$ \\
Replay buffer ($\left|\mathcal{D}\right|$) & $1{,}000$ & Decay factor ($\kappa$) & $0.995$ \\
\hline
\end{tabular}
\end{table}

\begin{table}[t]
\caption{Impact of hyperparameter choices on averaged classification accuracy (\%) of HDRL-MoE with \(K=4\) GDs, \(U=2\), and \(d_{\text{dev}}=0.8\) km. In each row, only the listed hyperparameter is changed while all other hyperparameters remain at their default values. Values in parentheses denote the performance difference compared to the default setting.}
\label{tab:hyper_sensitivity}
\centering
\renewcommand{\arraystretch}{1.15}
\setlength{\tabcolsep}{2.5pt}
\footnotesize
\resizebox{\linewidth}{!}{%
\begin{tabular}{c|c|c||c|c|c}
\hline
\multicolumn{3}{c||}{\textbf{Part I}} & \multicolumn{3}{c}{\textbf{Part II}} \\
\hline
\textbf{Hyperparam.} & \textbf{Value} & \textbf{Acc. (\%)} &
\textbf{Hyperparam.} & \textbf{Value} & \textbf{Acc. (\%)} \\
\hline
\multirow{3}{*}{\(\tau_{\xi}[0]\)} & \(0.3\) & 78.12 (-5.50) &
\multirow{3}{*}{\(\eta_{\text{c}}\)} & \(10^{-4}\) & 79.91 (-3.71) \\
 & \(1.0\) & 82.26 (-1.36) &
 & \(5\!\times\!10^{-4}\) & 82.41 (-1.21) \\
 & \(3.0\) & 76.55 (-7.07) &
 & \(10^{-2}\) & 81.64 (-1.98) \\
\hline
\multirow{3}{*}{\(\tau_{\omega}[0]\)} & \(0.3\) & 78.88 (-4.74) &
\multirow{3}{*}{\(\sigma_z^2[0]\)} & \(0.01\) & 80.31 (-3.31) \\
 & \(1.0\) & 82.71 (-0.91) &
 & \(0.05\) & 82.88 (-0.74) \\
 & \(3.0\) & 77.33 (-6.29) &
 & \(0.5\) & 82.22 (-1.40) \\
\hline
\multirow{3}{*}{\(\eta_{\text{a}}\)} & \(10^{-5}\) & 79.02 (-4.60) &
\multirow{3}{*}{\(\kappa\)} & \(0.95\) & 81.75 (-1.87) \\
 & \(5\!\times\!10^{-4}\) & 81.95 (-1.67) &
 & \(0.99\) & 83.42 (-0.20) \\
 & \(10^{-3}\) & 80.23 (-3.39) &
 & \(0.999\) & 82.96 (-0.66) \\
\hline

\hline
\rowcolor{gray!15}
\multicolumn{6}{c}{Default: all values in Table~\ref{tab:hyperparameters}; accuracy \(=83.62\)} \\
\hline
\end{tabular}
}
\end{table}

In this subsection, we present the hyperparameter settings for training the proposed \textbf{HDRL-MoE} framework and conduct a sensitivity analysis to investigate their impacts on system performance. We consider an independent and identically distributed (IID) data setting, where the STL-10 samples are randomly shuffled and evenly partitioned across GDs. The default hyperparameter settings used in our experiments are summarized in Table~\ref{tab:hyperparameters}. The actor and critic networks are trained using the Adam optimizer over $M=5000$ episodes. The sensitivity analysis is showcased in Table~\ref{tab:hyper_sensitivity}, which reveals the following crucial findings:
\begin{itemize}
    \item \textit{Gumbel-Softmax temperatures $\tau_{\xi}[0]$ and $\tau_{\omega}[0]$}: These are the most critical hyperparameters and exhibit the largest performance sensitivity across all evaluated settings. A low temperature, e.g., $0.3$, prematurely sharpens the action distributions, i.e., $\hat{\bm{z}}_{\xi}[m]$ and $\hat{\bm{z}}_{\omega,k}[m]$, which suppresses exploration and leads to severe accuracy drops. In contrast, a too large temperature value, e.g., $3.0$, makes the distributions nearly uniform, and the DRL agent is tempted to make random decisions.

    \item \textit{Learning rates $\eta_{\text{a}}$ and $\eta_{\text{c}}$}: Both learning rates govern a trade-off between convergence speed and training stability. For instance, decreasing $\eta_{\text{a}}$ to $10^{-5}$ reduces accuracy by $4.60\%$, while increasing it to $5\!\times\!10^{-4}$ and $10^{-3}$ causes accuracy drops of $1.67\%$ and $3.39\%$, respectively. This is because a small actor learning rate updates the offloading and trajectory policies inadequately within the finite training epochs, whereas an excessively large value can over-adjust the policy according to noisy critic estimates, resulting in less effective decisions.

    \item \textit{UAV actor noise variance $\sigma_z^2[0]$}: The exploration noise also plays an important role in training. On one hand, an insufficient variance, e.g., $0.01$, limits trajectory diversity and causes the UAV to converge to a locally suboptimal path. On the other hand, excessive noise, e.g., $0.5$ may introduce instability to the heading angle decisions.

    \item \textit{Decay factor $\kappa$}: This hyperparameter has the least influence on performance, with accuracy variations of at most $1.87\%$ across the evaluated range. A small value anneals the exploration noise too rapidly, causing premature convergence. However, an excessively large value maintains high exploration for too long, hindering the agent from exploiting the learned strategy in later training stages.
\end{itemize}

Overall, the sensitivity analysis shows that hyperparameter selection has a significant influence on the performance of \textbf{HDRL-MoE}, with the Gumbel-Softmax temperatures and learning rates being particularly important. Based on the results, we use the hyperparameter values in Table~\ref{tab:hyperparameters} by default as they provide the highest classification accuracy among the evaluated settings.

\begin{figure}[t]
    \centering
    \includegraphics[width=0.70\linewidth]{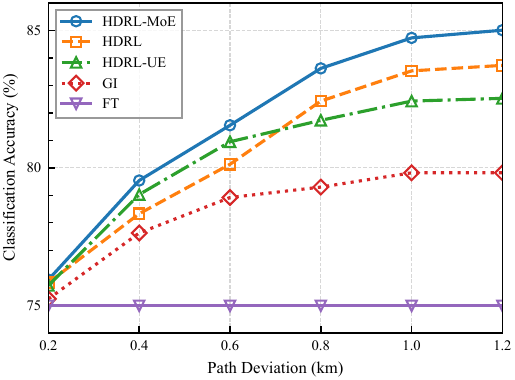}
    \caption{Classification accuracy (\%) versus path deviation requirement with $K=4$ GDs and $U=2$.}
    \label{fig:accuracy_vs_deviation}
\end{figure}

\subsection{Performance Evaluation}
To evaluate the effectiveness of the proposed \textbf{HDRL-MoE}, we conduct experiments under varying path deviation constraints and compare its classification accuracy performance against the baseline approaches. As shown in Fig.~\ref{fig:accuracy_vs_deviation}, \textbf{HDRL-MoE} consistently achieves higher classification accuracy than all baselines across various path deviation requirements. Notably, the performance across all methods is similar when the path deviation constraint is stringent, i.e., $d_{\text{dev}} \leq 0.4$ km, as mobility is severely restricted and the UAV cannot approach GDs for enhanced channel links to facilitate efficient inference task offloading. Moreover, \textbf{FT} exhibits nearly invariant accuracy with respect to $d_{\text{dev}}(\mathbf{Q})$, as the predetermined trajectory precludes it from adapting to improved channel conditions.
The greedy heuristic in \textbf{GI}, which makes offloading decisions based on the instantaneous uplink channel rates, fails to account for the future UAV trajectory and the long-term impact of mobility on system performance.
Therefore, the performance gain of the proposed \textbf{HDRL-MoE} over \textbf{GI} becomes more evident as the allowed path deviation $d_{\text{dev}}$ is relaxed. For instance, increasing $d_{\text{dev}}$ from $0.6$ km to $0.8$ km widens the performance gap between \textbf{HDRL-MoE} and \textbf{GI} from $2.63\%$ to $4.32\%$, respectively. Compared with \textbf{HDRL}, \textbf{HDRL-MoE} achieves up to $1.28\%$ higher classification accuracy, demonstrating the benefit of the MoE-based actor network architecture for learning the offloading policy. Furthermore, the accuracy improvements achieved by \textbf{HDRL-MoE} compared to \textbf{HDRL-UE} expands significantly, from $0.60\%$ to $2.30\%$, as the path deviation constraint is relaxed from $0.6$ km to $1.0$ km. This divergence arises as \textbf{HDRL-UE} is unable to effectively identify and prioritize GDs with more challenging data due to the absence of local prediction uncertainty in its state representation. In contrast, \textbf{HDRL-MoE} leverages the logit entropy to selectively offload data that are difficult for the local classifiers at GDs, while dynamically compressing intermediate feature to balance communication cost and offloading feature fidelity.

\begin{table}[t]
\centering
\caption{Successful offloading ratio (\%) for each GD with $d_{\text{dev}}=0.8$ km and $U=2$ over 200 test episodes on STL-10.}
\label{tab:offload}
\begin{tabular}{>{\columncolor{gray!15}}c|cccc|>{\columncolor{gray!15}}c}
\hline
\textbf{Methods} & {GD $1$} & {GD $2$} & {GD $3$} & {GD $4$} & {Averaged Value} \\
\hline
HDRL-MoE & 98.79 & 99.12 & 98.67 & 97.75 & 98.58 \\
HDRL-UE & 98.72 & 98.93 & 98.41 & 97.49 & 98.39 ($\downarrow$0.19) \\
FT & 98.98 & 99.32 & 98.45 & 97.96  & 98.63 ($\uparrow$0.05) \\
HDRL & 98.69 & 98.10 & 97.99 & 96.75 & 97.88 ($\downarrow$0.70) \\
GI & 94.44 & 97.08 & 95.13 & 95.83 & 95.62 ($\downarrow$2.96) \\
\hline
\end{tabular}
\end{table}

In UAV-assisted cooperative edge inference system, UAV-side inference improves performance only when intermediate features are successfully offloaded. Otherwise, GDs degenerate to local inference. Therefore, the successful offloading ratio is a key indicator of algorithm effectiveness, as it captures how a policy promotes cooperative inference. Table~\ref{tab:offload} summarizes the successful offloading ratios achieved by different algorithms under $d_{\text{dev}} = 0.8$ km. Notably, the average successful offloading ratios of \textbf{HDRL-MoE}, \textbf{HDRL-UE}, and \textbf{FT} are very close. However, the accuracy performance of \textbf{FT} is much lower because its fixed trajectory forces GDs to adopt more aggressive feature compression under weaker uplink channels. 
\textbf{HDRL-UE} achieves inferior performance despite a similar successful offloading ratio, since it cannot effectively identify samples that benefit most from UAV-assisted inference, as will be further discussed in the next subsection. Besides, \textbf{HDRL} and \textbf{GI} exhibit lower average successful offloading ratios of $97.88\%$ and $95.62\%$, respectively, resulting in performance degradations of $0.70\%$ and $2.96\%$ compared to \textbf{HDRL-MoE}. This outcome indicates the limitations of \textbf{HDRL} in learning effective inference decisions without the MoE-based architecture. In addition, the greedy heuristic offloading policy in \textbf{GI} fails to consider future UAV trajectory and long-term system performance, thereby leading to reduced successful offloading ratios. 

\subsection{Analysis under Non-IID Data Distributions}
\begin{figure}[t]
    \centering
    \begin{subfigure}[t]{0.42\linewidth}
        \centering
        \includegraphics[width=\linewidth]{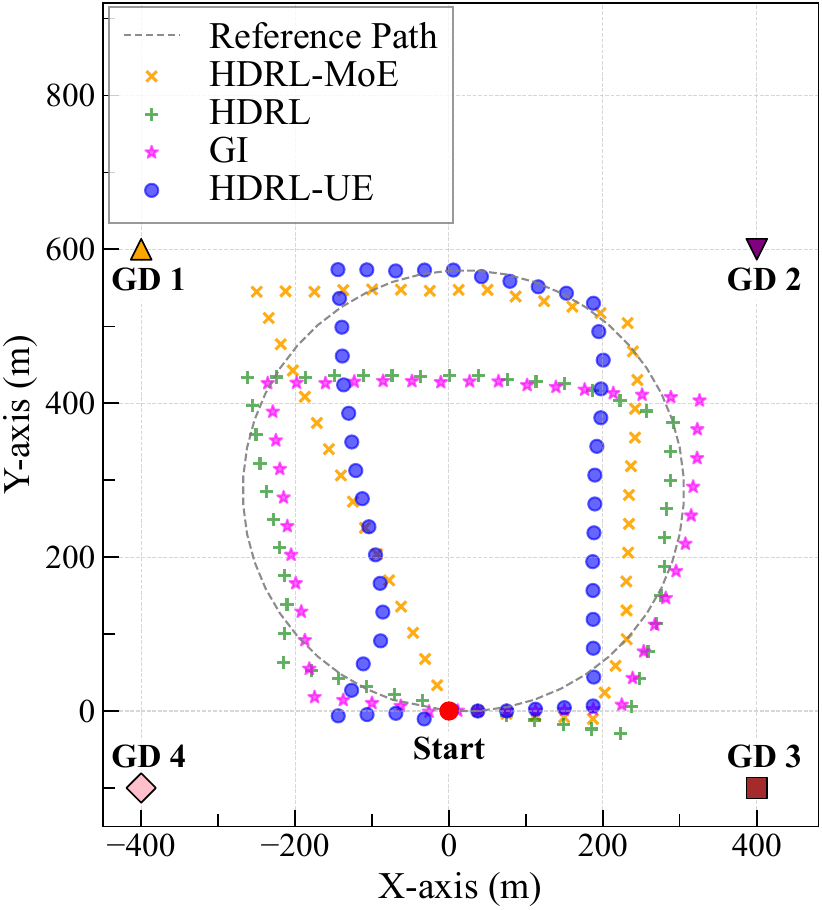}
        \caption{}
        \label{fig:non-iid-traj}
    \end{subfigure}
    \hfill
    \begin{subfigure}[t]{0.45\linewidth}
        \centering
        \raisebox{0.4cm}{\includegraphics[width=\linewidth]{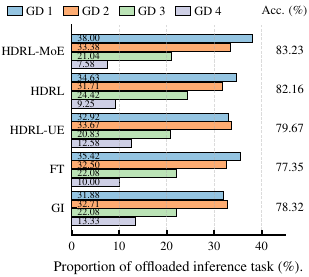}}
        \caption{}
        \label{fig:non-iid-acc}
    \end{subfigure}
    \caption{(a) UAV trajectory of different algorithms under non-IID data distributions. (b) Proportion of offloaded inference task (\%) from each GD and average classification accuracy (\%) with $d_{\text{dev}}=0.8$ km and $U=2$. The results are averaged over 200 test episodes on STL-10.}
    \label{fig:non-iid}
\end{figure}

We further evaluate the proposed algorithm and the baselines under the scenarios where each GD holds a non-IID data distribution. Specifically, we synthesize non-IID data across $K=4$ GDs by partitioning the STL-10 dataset based on the hardness of samples, measured via the prediction entropy of the on-device model. The dataset is first sorted in descending order of entropy, then evenly divided into $K$ subsets, with GD $1$ receiving the hardest samples and GD $4$ the easiest ones. As illustrated in Fig.~\ref{fig:non-iid}(a), the UAV trajectories learned by different algorithms exhibit distinct patterns under non-IID data distributions. Specifically, \textbf{HDRL-MoE} prioritizes flying closer to GDs with more challenging data, i.e., GD $1$ and GD $2$, to enhance their offloading success rates and improve overall inference accuracy. However, \textbf{GI} favors GD $3$ and GD $4$ that possess easier samples, resulting in suboptimal classification performance. This inefficiency stems from its inability to forecast channel capacity variations over the UAV trajectory, leading to misguided offloading decisions. Consequently, critical samples from GD $1$ and GD $2$ often fail to be offloaded successfully due to the insufficient link quality.

\begin{table}[t]
\centering
\caption{Classification Accuracy (\%) versus number of GDs with $d_{\text{dev}} = 0.8$ km and $U = K/2$.}
\label{tab:scale}
\begin{tabular}{c|cccc}
\hline
\textbf{Method} & \textbf{$K=2$} & \textbf{$K=4$} & \textbf{$K=8$} & \textbf{$K=16$} \\
\hline
HDRL-MoE & 82.03 & 81.55 & 80.60 & 79.23 \\
HDRL & 81.88 & 80.13 & 77.19 & 74.53 \\
\hline
\rowcolor{gray!15} Performance Gain & 0.15 & 1.42 & 3.41 & 4.70 \\
\hline
\end{tabular}
\end{table}

We also quantify the offloading decisions and classification accuracy of different algorithms under non-IID data distributions, which are summarized in Fig.~\ref{fig:non-iid}(b). While \textbf{HDRL} and \textbf{HDRL-UE} similarly attempt to navigate toward GD $1$ and GD $2$, it reveals that their learned trajectories remain suboptimal as fewer offloading opportunities are allocated to these GDs compared to \textbf{HDRL-MoE}. 
Specifically, \textbf{HDRL-MoE} outperforms \textbf{HDRL} by $1.07\%$ in classification accuracy, owing to the ability of the MoE-based architecture to learn a more effective task offloading policy that prioritizes GDs with more challenging data.
The task offloading policy is then synergistically integrated with UAV trajectory optimization via the hierarchical network architecture, further enhancing the offloading success rates of critical samples from GD $1$ and GD $2$.
Moreover, \textbf{HDRL-UE} suffers a significant accuracy drop of $3.56\%$ compared to \textbf{HDRL-MoE}, highlighting the critical role of local prediction uncertainty in guiding offloading decisions under non-IID data distributions. 
Similar to the IID data scenario, \textbf{FT} exhibits lower performance than \textbf{HDRL-MoE} under non-IID data due to its non-adaptive fixed trajectory.

\begin{figure}[t]
    \vspace{-0.5em}
    \centering
    \begin{subfigure}[t]{0.40\linewidth}
        \centering
        \includegraphics[width=0.90\linewidth]{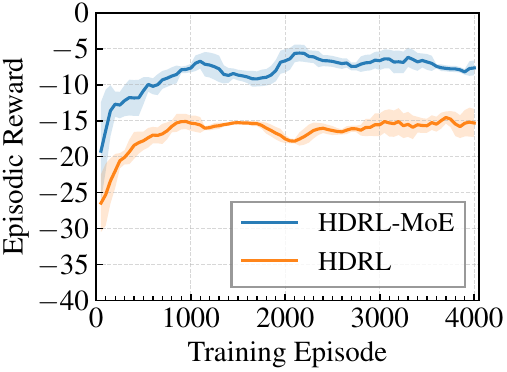}
        \caption{$K=2$}
    \end{subfigure}
    \begin{subfigure}[t]{0.40\linewidth}
        \centering
        \includegraphics[width=0.90\linewidth]{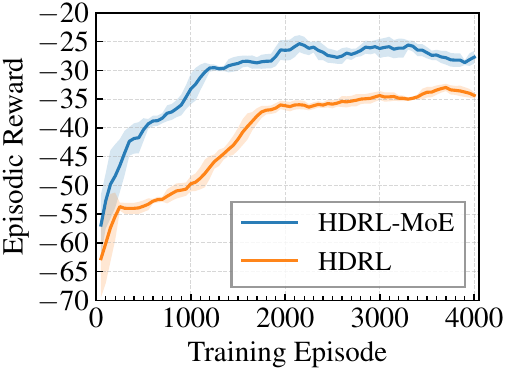}
        \caption{$K=4$}
    \end{subfigure}
    \begin{subfigure}[t]{0.40\linewidth}
        \centering
        \includegraphics[width=0.90\linewidth]{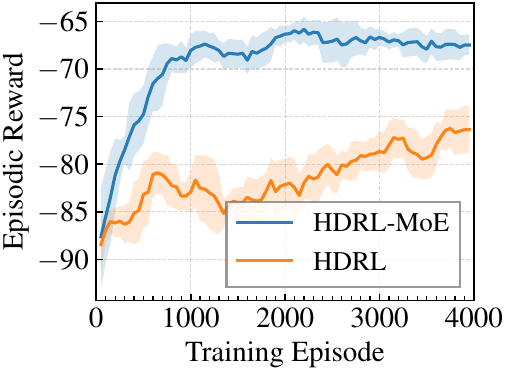}
        \caption{$K=8$}
    \end{subfigure}
    \begin{subfigure}[t]{0.40\linewidth}
        \centering
        \includegraphics[width=0.90\linewidth]{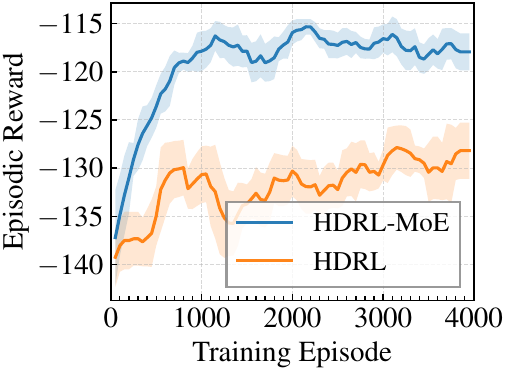}
        \caption{$K=16$}
    \end{subfigure}
    \caption{Training reward convergence of \textbf{HDRL-MoE} versus \textbf{HDRL} with $d_{\text{dev}}=0.8$~km, $K \in \{2,4,8,16\}$ GDs, and $U=K/2$. Each point is the mean reward over a 5-episode sliding window with shaded regions showing the standard deviation.}
    \label{fig:scale}
\end{figure}

\subsection{Scalability to Multiple GDs}
In this part, we compare \textbf{HDRL-MoE} against \textbf{HDRL} as the number of GDs varies to assess the scalability of the MoE-based cooperative edge inference architecture. We evaluate \textbf{HDRL-MoE} with \(K \in \{2,4,8,16\}\) GDs at a fixed path-deviation constraint \(d_{\text{dev}}=0.8\) km.
As shown in Table~\ref{tab:scale}, \textbf{HDRL-MoE} maintains higher classification accuracy than \textbf{HDRL} across the tested system scales and exhibits only modest degradation as $K$ increases. Notably, the performance gap between the two methods becomes more prominent with more GDs. For instance, the absolute accuracy gap between the two methods rises from $0.15\%$ at $K=2$ to $4.70\%$ at $K=16$. This is because the joint action space of offloading decisions and compression ratios grows exponentially with $K$, making it increasingly difficult for \textbf{HDRL} to learn an effective policy. In contrast, the proposed MoE-based architecture decomposes the global action space with the router and expert networks. Therefore, the actor network complexity of the MoE-based architecture grows linearly as $K$ increases, which substantially mitigates the curse of dimensionality. 
Moreover, in cases with larger numbers of GDs i.e., $K=8$ and $K=16$, \textbf{HDRL-MoE} reaches its highest reward within 2500 episodes, whereas \textbf{HDRL} requires approximately 3500 episodes. The reduction in training episodes demonstrates the learning efficiency of the proposed framework.

\section{Conclusions}
This paper investigated a UAV-assisted cooperative edge inference framework for low-altitude economy scenarios, where a UAV simultaneously executes mission-critical LAE duties and supports ground devices (GDs) in AI inference tasks by offloading intermediate features. The joint optimization of UAV trajectories and offloading policies was formulated as a constrained partially observable Markov decision process, which presents multifaceted challenges due to the coupling between UAV mobility and inference decisions, and the combinatorial nature of offloading decisions and compression ratios across multiple GDs. To address these challenges, we propose HDRL-MoE, a hierarchical deep reinforcement learning algorithm that employs a two-timescale structure to decouple trajectory and inference decisions, and a mixture-of-experts actor network to efficiently handle the combinatorial action space. Extensive simulations demonstrated that the proposed algorithm significantly improves inference accuracy compared with baseline schemes, as well as its superior scalability to more GD scenarios. Future work will extend the framework to multi-UAV coordination and more realistic airspace regulatory constraints.

\bibliographystyle{IEEEtran}
\bibliography{ref}

\end{document}